\documentclass[superscriptaddress,twocolumn,showpacs,amsmath,amssymb, prb, showkeys]{revtex4-1}
\usepackage{amsmath}
\usepackage{stmaryrd}
\usepackage{txfonts}
\usepackage{amssymb}
\usepackage{mathrsfs}
\usepackage{graphicx}
\usepackage{dcolumn}
\usepackage{bm}
\usepackage{epsfig}
\usepackage{color}

\usepackage[colorlinks=true, linkcolor=blue, citecolor=blue, urlcolor=blue]{hyperref}
\begin{document}

\title{Realization of the Axial Next-Nearest-Neighbor Ising model in U$_3$Al$_2$Ge$_3$}

\author{David M. Fobes}
\affiliation{MPA-CMMS, Los Alamos National Laboratory, Los Alamos, New Mexico 87545, USA}

\author{Shi-Zeng Lin}
\affiliation{Theoretical Division, T-4 and CNLS, Los Alamos National Laboratory, Los Alamos, New Mexico 87545, USA}

\author{Nirmal J. Ghimire}
\affiliation{MPA-CMMS, Los Alamos National Laboratory, Los Alamos, New Mexico 87545, USA}
\affiliation{Argonne National Laboratory, Lemont, Illinois 60439, USA}

\author{Eric D. Bauer}
\affiliation{MPA-CMMS, Los Alamos National Laboratory, Los Alamos, New Mexico 87545, USA}

\author{Joe D. Thompson}
\affiliation{MPA-CMMS, Los Alamos National Laboratory, Los Alamos, New Mexico 87545, USA}

\author{Markus Bleuel}
\affiliation{NIST Center for Neutron Research, Gaithersburg, Maryland 20899, USA}
\affiliation{Department of Materials Science and Engineering, University of Maryland, College Park, MD 20742-2115, USA}

\author{Lisa M. DeBeer-Schmitt}
\affiliation{CEMD, Oak Ridge National Laboratory, Oak Ridge, Tennessee 37831, USA}

\author{Marc Janoschek}
\affiliation{MPA-CMMS, Los Alamos National Laboratory, Los Alamos, New Mexico 87545, USA}
\email{mjanoschek@lanl.gov}

\begin{abstract}
Here we report small-angle neutron scattering (SANS) measurements and theoretical modeling of U$_3$Al$_2$Ge$_3$. Analysis of the SANS data reveals a phase transition to sinusoidally modulated magnetic order, at $T_{\mathrm{N}}=63$~K to be second order, and a first order phase transition to ferromagnetic order at $T_{\mathrm{c}}=48$~K. Within the sinusoidally modulated magnetic phase ($T_{\mathrm{c}} < T < T_{\mathrm{N}}$), we uncover a dramatic change, by a factor of three, in the ordering wave-vector as a function of temperature. These observations all indicate that U$_3$Al$_2$Ge$_3$ is a close realization of the three-dimensional Axial Next-Nearest-Neighbor Ising model, a prototypical framework for describing commensurate to incommensurate phase transitions in frustrated magnets.
\end{abstract}

\date{\today}
\maketitle

\section{Introduction}
The Axial Next-Nearest-Neighbor Ising (ANNNI) model is a historical framework which has successfully described frustrated magnetism in a variety of materials, and in particular, commensurate to incommensurate phase transitions.\cite{bak_commensurate_1982,selke_annni_1988} The three-dimensional ANNNI model, describing the competing nearest neighbor (NN) interaction $J_1$, and the next nearest neighbor (NNN) interaction $J_2$, between Ising moments in one direction and simple ferromagnetic (FM) interactions between Ising moments in the same plane, was first proposed by R.\ J.\ Elliott in 1961,\cite{PhysRev.124.346} and has been studied extensively ever since. Subsequently, the ANNNI model was generalized to lower dimensions, and to $XY$ or Heisenberg spins.\cite{selke_annni_1988,yamashita_1998,harris_1990} This model and its variants have been shown to be relevant to a broad class of systems, including alloys, magnets, ferroelectrics and adsorbates.\cite{bak_commensurate_1982,selke_annni_1988}

For the case of magnetic materials, compounds containing rare-earth or actinide ions with $4f$ and $5f$ electrons, respectively, are expected to feature ANNNI physics due to the presence of Ruderman-Kittel-Kasuya-Yosida (RKKY) interactions, which can be approximated in many cases by the competing NN and NNN exchange integrals.\cite{Coleman2010} $f$-electron materials, where the Fermi-surface-topology-determined RKKY interaction also competes with crystal field effects and spin orbit coupling, contain rich phase diagrams with a multitude of magnetically ordered states. For example, in the presence of a quasi-nested Fermi surface, the RKKY interaction becomes maximal at a nonzero wave vector $\mathbf{Q}$, resulting in a magnetic spiral ground state, with a period $2\pi/Q$, typically incommensurate with the underlying chemical lattice.\cite{jensen_1991,das_2015,fobes_jpcm_2017} With the addition of uniaxial magnetic anisotropy, the magnetic spiral can become distorted, resulting in higher harmonic wave vectors, and potential quasi-continuous changes to the ordering wave vector $\mathbf{Q}$ as a function of temperature, arising due to the competition of RKKY interactions and thermal fluctuations.\cite{zaliznyak_1995} Although the ANNNI model provides a theoretical framework to describe this succession of long-period incommensurate phases as function of temperature, it was found that close fulfillment of its predictions for the temperature dependence of $\mathbf{Q}$ is rare. Maybe the best known realization of the ANNNI model in $f$-electron materials is the semimetal CeSb,\cite{fischer_1978,meier_1978,pokrovskii_1982} with the caveat that additional further-neighbor interactions were found to be important.\cite{meier_1978,pokrovskii_1982}

In this work, we demonstrate that the tetragonal $f$-electron compound U$_3$Al$_2$Ge$_3$ is a close realization of the three-dimensional ANNNI model. Hallmarks of the ANNNI model include (i) a magnetic phase which features a temperature-dependent magnetic ordering wave vector, owing to a competition between the NN and NNN interactions and magnetic fluctuations, in the vicinity of the phase transition to the paramagnetic (PM) phase, between which lies a second order phase transition, and (ii) a FM phase at low temperature that is entered via a first order transition, at which the magnetic ordering wave vector features a logarithmic singularity.\cite{selke_annni_1988}  A previous neutron powder diffraction study suggests that U$_3$Al$_2$Ge$_3$ features several of these ingredients, including a low temperature FM phase below $T_{\mathrm{c}}=48~\mathrm{K}$, an incommensurately sinusoidally modulated magnetic state ($T_{\mathrm{c}} < T < T_{\mathrm{N}}$, with $T_{\mathrm{N}}=63~\mathrm{K}$) with ordering wave vector $\mathbf{Q}=[0,0,\delta]$, and a second-order transition between said phase and a PM phase, making this material a potential candidate for ANNNI realization.\cite{rogl_magnetic_1999} However, to confirm if  U$_3$Al$_2$Ge$_3$ is indeed well-characterized by the ANNNI model several key aspects need to be clarified. (a) Although the incommensurate magnetic phase was reported to be sinusoidally modulated, symmetry, in principle, also allows for spiral magnetic order, which would not agree with the ANNNI model. Notably, powder diffraction can frequently not unambiguously distinguish between a sinusoidal and spiral magnetic order. (b) Previous to our study, no temperature dependence of the magnetic ordering wave vector has been reported. (c) The nature of the transition between the FM and incommensurate phases remains unknown.

Here, to address these issues, we have carried out a detailed neutron scattering study on a single crystal of U$_3$Al$_2$Ge$_3$ in combination with theoretical modeling based on the ANNNI model. Using small-angle neutron scattering (SANS), a powerful technique for accurately studying long-wavelength magnetic structures, we have confirmed that $\mathbf{Q}=[0,0,\delta]$ indeed shows a pronounced temperature dependence, and determined the nature of the magnetic phase transitions. Our results clearly reveal a second order phase transition from paramagnetic to sinusoidally modulated Ising moments at $T_{\mathrm{N}}=63$~K, followed by a first-order phase transition at $T_{\mathrm{c}}=48$~K to a FM state. Through theoretical modeling we are able to ascribe these experimental observations to the ANNNI framework; both the measured and calculated phase diagram can be understood in terms of the 3D ANNNI lattice model, with frustrated interactions in the direction of modulation, given by the ratio $J_2/J_1=-0.2815$, where $J_1$ is the nearest neighbor ferromagnetic interaction, $J_2$ is the next nearest neighbor antiferromagnetic interaction,\cite{selke_annni_1988} and the ratio $J_2/J_1$ is given by the $Q$ vector at $T_{\mathrm{N}}$, \textit{i.e.} $\cos(\delta c/ 2\pi)=-J_1/4J_2$, in which $c$ is the $c$-axis lattice parameter. Furthermore, at the incommensurate-to-FM phase transition we confirm the presence of a logarithmic singularity in the temperature dependence of $\mathbf{Q}$, an essential feature of the three-dimensional ANNNI model, as stated above. Therefore, we find that U$_3$Al$_2$Ge$_3$ is a close realization of ANNNI model, offering an important playground for investigating its rich physics. 

The manuscript is organized in the following way: In Section \ref{model} we describe the theoretical model that motivated our experimental study. In Sections \ref{synthesis} and \ref{results} we report the sample synthesis and characterization, and our SANS results, respectively. Finally, in Section \ref{discussion} we will discuss and summarize our combined theoretical and experimental results.

\section{Model}
\label{model}
To motivate this experimental study we develop the following model accounting for the experimentally observed magnetic phase diagram \cite{rogl_magnetic_1999}, making several experimental predictions. The neutron powder diffraction results\cite{rogl_magnetic_1999} determined that in U$_3$Al$_2$Ge$_3$ the magnetic moments lying on the Uranium $8c$ and $2a_1$ symmetry sites (but not the $2a_2$ sites), with magnetic moments $\mu_U=2.37\ \mu_B$ and $\mu_U=2.12\ \mu_B$, respectively, align parallel to the $a$-axis due to strong easy-axis anisotropy, modulate along the $c$-axis, and order ferromagnetically in the $ab$-plane, allowing us to reduce the problem to one dimension. Due to the small ordering wave vector, we expand the magnetic free energy in terms of the magnetization density $\mathbf{M}$ and ordering wave vector $\mathbf{Q}$ in the continuum limit,
\begin{align}\label{eq1}
\begin{split}
{\cal F}({\mathbf{M}}) = \frac{\alpha }{2}\mathbf{M}^2 + \frac{\beta }{4}\mathbf{M}^4 + \frac{\delta }{6}\mathbf{M}^6 - \frac{\mu }{2}{(\partial_z {\mathbf{M}})^2} \\+ \frac{\eta }{2}{({\partial_z ^2}{\mathbf{M}})^2} 
+ \frac{\gamma }{2}\mathbf{M}^2{(\partial_z {\mathbf{M}})^2} - \frac{{{A_2}}}{2}M_x^2,
\end{split}
\end{align}
where $A_2$ is the easy-axis anisotropy, and $\mu$, $\eta$ and $\gamma$ are associated with the stiffness of the magnetic modulation. The competing interactions $J_1$ and $J_2$ along the $c$-axis in the ANNNI lattice model are captured by the $\mu>0$ and $\eta>0$ terms. Approaching the N\'{e}el temperature $T_{\mathrm{N}}$ from the paramagnetic phase, we can neglect the quartic and higher order terms of $\mathbf{M}$. To take advantage of the anisotropy,  the system stabilizes a state with sinusoidal moment modulation, $M_x\propto\sin(Q r)$ and $M_y=M_z=0$. We obtain the optimal $Q$ by minimizing $\mathcal{F}$ with respect to $Q$, which yields $Q_o=\sqrt{\mu/2\eta}$ at $T_{\mathrm{N}}$. Here $T_{\mathrm{N}}$ is determined by the condition that the coefficient of the quadratic term in $M_x$ vanishes 
\begin{align}
{\alpha } - {{{A_2}}} - \frac{{{\mu ^2}}}{{4\eta }} = 0,
\end{align}
which is enhanced due to the presence of the easy-axis anisotropy $A_2$. To fulfill the condition for second order phase transition between incommensurate and paramagnetic phase at $T_{\mathrm{N}}$, we assume $\beta>0$.

Lowering the temperature results in an increase of the magnetic moment. Higher-order terms and easy-axis anisotropy both tend to distort the simple sinusoidal modulation. The wave is squared up as moments are forced to align along the easy-axis, creating harmonics in the wave vectors. To take the advantage of the anisotropy, the moments lie in the easy-axis direction. The magnetic state can be described by an elliptic function
\begin{align}\label{eq3}
{M_x} = \Delta\  \mathrm{sn}({z}/{\xi },k),\ \ M_y=M_z=0,
\end{align}
where $\mathrm{sn}(z)$ is the Jacobi elliptic function. The meanings of $\xi$, $k$, $\Delta$ become clear in the limit $k\rightarrow 0$, $\mathrm{sn}(z, k=0)=\sin(z)$: $k$ describes the deviation from a perfect sinusoidal modulation, thus reflecting the importance of harmonics for the magnetic modulation, $\xi$ is a length scale appearing in the period of the magnetic modulation, and $\Delta$ is the amplitude of the modulation.
When the coefficients in Eq. \eqref{eq1} satisfy a certain relation (shown below), the magnetic state can be found exactly using Eq. \eqref{eq3}. For generic parameters, we can use variational calculations using Eq. \eqref{eq3} to determine $\xi$, $k$ and $\Delta$.

\begin{figure}[t]
\includegraphics[width=1\columnwidth]{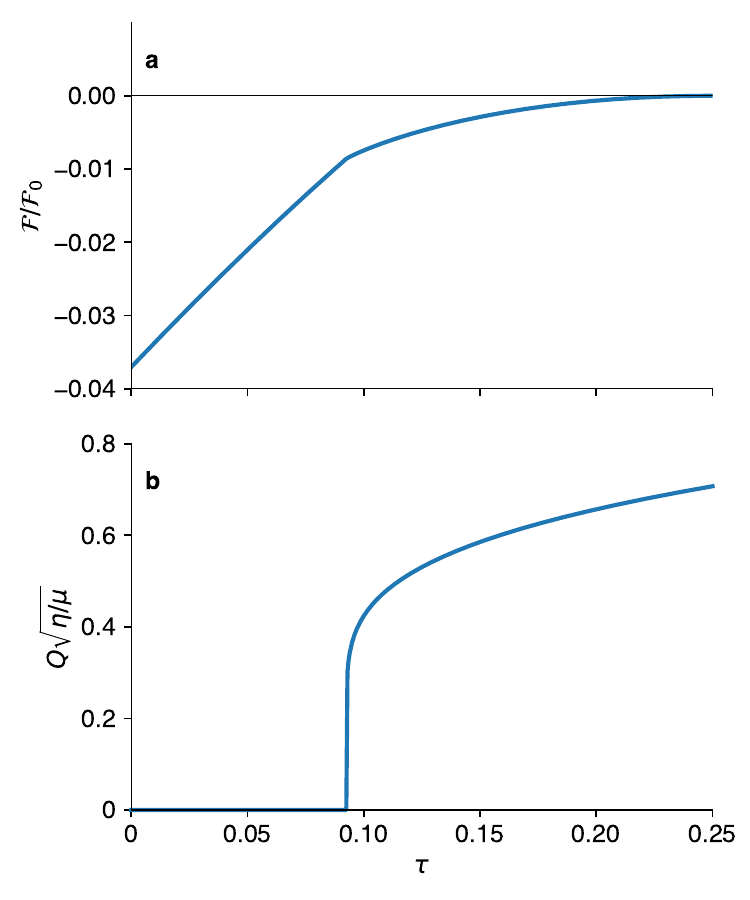}
\caption{(color online) Temperature dependence of (a) free energy $\mathcal{F}$, and (b) wave vector $Q$ for model in Eq. \eqref{eq1}. Near the commensurate-incommensurate transition, $Q$ follows Eq. \eqref{eq14}. The magnetic order sets in at $\tau=1/4$. We have used $\beta\gamma/\delta\mu=-10/3$ in the calculations.
} \label{f3}
\end{figure}

Let us first construct the exact solution.\cite{buzdin_1997} We note that $M_x$ in Eq. \eqref{eq3} solves the following equations  
\begin{align}\label{eq4}
{\xi ^2}\partial _z^2{M}_x + (1 + {k^2}){M}_x - 2\frac{{{k^2}}}{{{\Delta ^2}}}M_x^3 = 0,
\end{align} 
\begin{align}\label{eq5}
{\xi ^2}{({\partial _z}{M_x})^2} + (1 + {k^2})M_x^2 - \frac{{{k^2}M_x^4}}{{{\Delta ^2}}} = {\Delta ^2}.
\end{align} 
Differentiating Eq. \eqref{eq4} with respect to $z$ twice, we obtain
\begin{align}\label{eq6}
\begin{split}
\partial _z^4{M_x} + \frac{{(1 + {k^2})}}{{{\xi ^2}}}\partial _z^2{M_x} + \frac{{{12k^4}}}{{{\Delta ^4}{\xi ^4}}}M_x^5 - \frac{{6{k^2}(1 + {k^2})}}{{{\Delta ^2}{\xi ^4}}}M_x^3 \\
- 12\frac{{{k^2}}}{{{\Delta ^2}{\xi ^2}}}[{M_x}{({\partial _z}{M_x})^2} + M_x^2\partial _z^2{M_x}]
 = 0.
\end{split}
\end{align}
Multiplying Eq. \eqref{eq4} by $M_x^2$ and Eq. \eqref{eq5} by $M_x$ and then adding them together, we have
\begin{align}\label{eq7}
\begin{split}
\frac{{{k^2}}}{{{\Delta ^2}{\xi ^2}}}\left[M_x^2\partial _z^2{M_x} + {M_x}{({\partial _z}{M_x})^2}\right] + \frac{{2{k^2}(1 + {k^2})}}{{{\Delta ^2}{\xi ^4}}}M_x^3 \\
- \frac{{{3k^4}}}{{{\Delta ^4}{\xi ^4}}}M_x^5 - \frac{{{k^2}}}{{{\xi ^4}}}{M_x} = 0.
\end{split}
\end{align}
We multiply Eq. \eqref{eq7} by a factor $A$ and Eq. \eqref{eq4} by a factor $B(1+k^2)/\xi^4$ and add to Eq. \eqref{eq6}. We then compare the resulting equation to Eq. \eqref{eq1}, from which we obtain the equations for coefficients
\begin{align}\label{eq8}
\begin{split}
(12 - A)\frac{{{k^2}}}{{{\Delta ^2}{\xi ^2}}} = \frac{\gamma }{\eta }, \\  (B + 1)\frac{{(1 + {k^2})}}{{{\xi ^2}}} = \frac{\mu }{\eta }, \\ \frac{{(12 - 3A){k^4}}}{{{\Delta ^4}{\xi ^4}}} = \frac{\delta }{\eta }, \\
\frac{{{k^2}(1 + {k^2})(2A - 6 - 2B)}}{{{\Delta ^2}{\xi ^4}}} = \frac{\beta }{\eta },\\ \frac{{{{(1 + {k^2})}^2}B - A{k^2}}}{{{\xi ^4}}} = \frac{{\alpha '}}{\eta },
\end{split}
\end{align}
with $\alpha'=\alpha-A_2$. In principle we can determine $A$, $B$, $k$, $\Delta$ and $\xi$ from Eq. \eqref{eq8}; however, the solutions are not guaranteed to exist for arbitrarily coefficients of $\mathcal{F}$, as can been seen from the equations with $\gamma$ and $\delta$. This generally implies that Eq. \eqref{eq3} is not an exact solution. We consider the case of a particular set of coefficients, such that Eq. \eqref{eq3} is the exact solution, to demonstrate explicitly the dependence of $M_x$ on temperature. In any case, Eq. \eqref{eq3} should be a good variation ansatz to describe the spatial profile of $M_x$.

At $T_{\mathrm{N}}$ corresponding to $\alpha'=\mu^2/4\eta$, the modulation of $M_x$ should be a sinusoidal wave. We have $\Delta=0$ and $k=0$ while $\Delta_k=\Delta/k=\mathrm{constant}$, in the limit $k\rightarrow 0$. We obtain $B=1$, $\xi^2=2\eta/\mu$, $\Delta_k=-3\beta/2\delta$. As in the usual Ginzburg-Landau theory, we assume that the temperature dependence can be modeled in $\alpha'$ by introducing $\tau  = \alpha '/(\frac{{{\mu ^2}}}{\eta })$, and other coefficients do not depend on $T$. Here $\tau$ is the reduced temperature. Upon a change in temperature, $k$ and $B$ change continuously, but both $A$ and $\Delta_k\xi$ are independent of $T$. We have
\[
B(\tau ,k) =  - \frac{{ - 1 + 2\tau  + \sqrt {1 - 4[1 + (12 + \frac{{3\beta \gamma }}{{\delta \mu }}){k^2}{{(1 + {k^2})}^{ - 2}}]\tau } }}{{2\tau }}.
\]
\[
\xi  = \sqrt {(B + 1)\frac{{(1 + {k^2})\eta }}{\mu }},\ \ {\Delta _k} = \sqrt {\frac{-3\beta\eta}{\delta(B+1)(1+k^2)}} 
\]
The corresponding free energy density
\begin{align}\label{eq9}
\frac{{{{\cal F}}}}{{{{\cal F}_0}}} = \frac{{ - {k^4}\left[10{I_2} - ({k^2} + 1)(11 + B){I_4} + 14{k^2}{I_6}\right]}}{{{{[({k^2} + 1)(B + 1)]}^3}}},
\end{align}
where
\begin{align}
\mathcal{F}_0=\frac{{-3\beta\mu^2 }}{{2\delta {\eta}}},\ \ \   {I_2} = \frac{1}{{{k^2}}}\left[1 - \frac{{E(k)}}{{K(k)}}\right],
\end{align}
\begin{align}
{I_4} = \frac{{2 + {k^2}}}{{3{k^4}}} - \frac{{2(1 + {k^2})}}{{3{k^4}}}\frac{{E(k)}}{{K(k)}},
\end{align}
\begin{align}
{I_6} = \frac{{4({k^2} + 1)}}{{5{k^2}}}{I_4} - \frac{3}{{5{k^4}}}\left[1 - \frac{{E(k)}}{{K(k)}}\right],
\end{align}
with $E(k)$ and $K(k)$ being the complete elliptic integral of the second and first kind, respectively. We have used the convention $E(k)\equiv\int_0^{\pi/2}\sqrt{1-k^2\sin^2\theta}d\theta$. To determine the optimal $k$, we numerically minimize the free energy density $\mathcal{F}$, given by Eq. \eqref{eq9}, with respect to $k$, and then obtain other parameters from $k$.  We then obtain $\xi$.

The period of the sinusoidal wave is given by the expression
\begin{align}\label{eq14}
{\lambda _I} = 4\xi K(k). 
\end{align}
Here $\lambda_I$ diverges when $k\rightarrow 1$, corresponding to the FM state. Near the phase transition between sinusoidal wave and FM at $\tau_2$, we can expand $K(k)$ near $k=1$,
\begin{align}\label{eq15}
{\lambda _I}=\xi(\tau_2)\left[8\ln 2-2 \ln(1-k^2)\right],
\end{align}
Close to $\tau_2$ from above, we expand $1-k^2=\alpha_2(\tau-\tau_2)^{\beta_2}$ and then obtain the logarithmic temperature dependence of the ordering wave vector
\begin{align}\label{eq15cc}
Q=2\pi/\lambda_I\propto -1/\ln(\tau-\tau_2),
\end{align}
near the commensurate-incommensurate (sinusoidal modulation-FM) transition. \cite{bak_commensurate_1982,selke_annni_1988}

The free energy $\mathcal{F}$ and $Q$ as a function of temperature $\tau$ are presented in Fig. \ref{f3}. Upon lowering the temperature from $T_{\mathrm{N}}$, the system first becomes an Ising density wave with $Q$ varying continuously with $T$, becoming FM at low temperature, with the free energy density
\begin{align}
\frac{{{{\cal F}_{FM}}}}{{{{\cal F}_0}}} =  - \frac{1}{{18}}\left[(1 + \sqrt {1 - 6\tau } )\left(\frac{1}{3} - 2\tau \right) - \tau \right].
\end{align}
The transition from the paramagnetic phase into the sinusoidally-modulated phase is of second order by construction, following the experiments shown below and Ref. \onlinecite{rogl_magnetic_1999}, and the transition from the sinusoidally-modulated phase to the FM is shown to be first order, which can be seen from the slope of $\mathcal{F}$ in Fig. \ref{f3} (a). As we will show, these results agree qualitatively with results of the experimental analysis of the SANS data (cf. Fig. \ref{f2}(b)).

\section{Synthesis and Characterization}
\label{synthesis}
\begin{figure}[t]
\includegraphics[width=1\columnwidth]{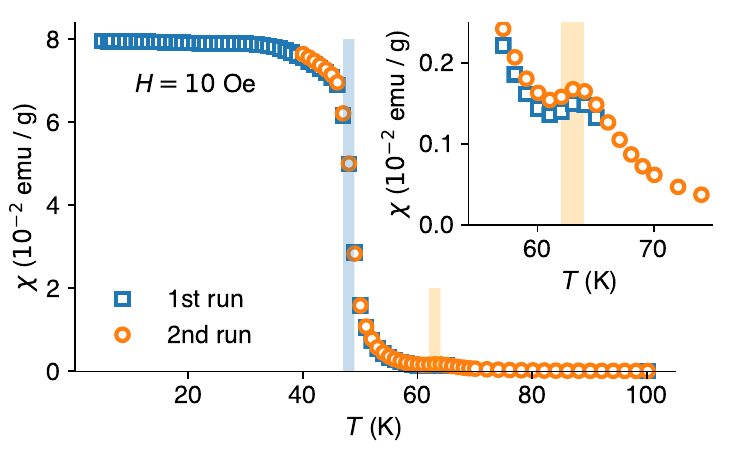}
\caption{(Color Online) Magnetic susceptibility $\chi$ of a single crystal of U$_3$Al$_2$Ge$_3$ taken with $H=10$~Oe. Vertical lines indicate second- and first-order transitions at $T_{\mathrm{N}}=63~\mathrm{K}$ and $T_{\mathrm{c}}=48~\mathrm{K}$, respectively. Inset: Zoomed region of magnetic susceptibility $\chi$ showing the second-order transition at $T_{\mathrm{N}}=63~\mathrm{K}$. Different symbols represent separate runs on the same sample.
} \label{f0}
\end{figure}

\begin{figure}[h!]
\includegraphics[width=0.99\columnwidth]{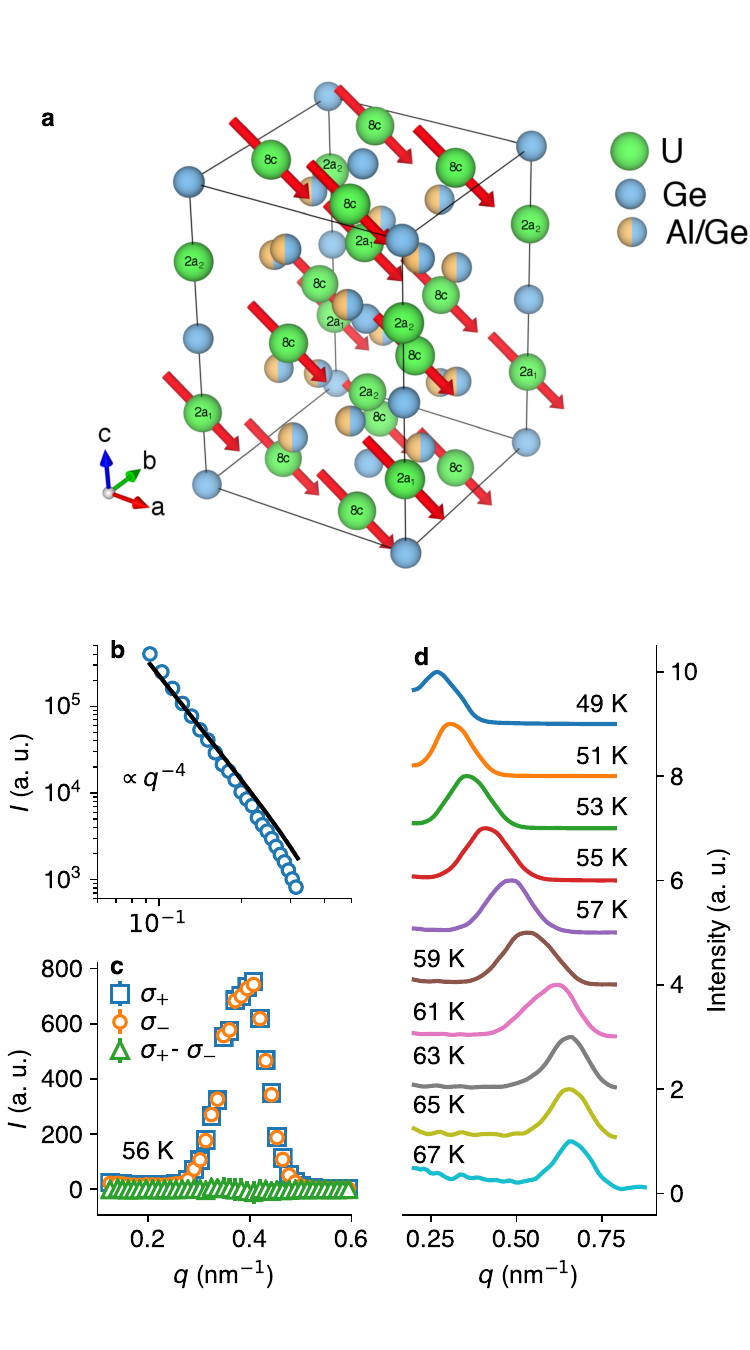}
\caption{(Color Online) (a) U$_3$Al$_2$Ge$_3$ tetragonal crystal structure and the magnetic configuration in the ferromagnetic phase, with the three U site symmetries individually labeled ($8c$, $2a_1$, $2a_2$). (b) Low-$q$ scattering in the ferromagnetic phase ($T=40$~K), fitted to Porod's Law $I \propto q^{-4}$, illustrating the expected behavior for FM domains. (c) Polarized SANS cross-sections with the incident beam polarized parallel ($\sigma_+$) and antiparallel ($\sigma_-$) to the momentum transfer $q$, and their difference, taken at NG7-SANS. As explained in the text, the difference should only be non-zero in the case of a spiral magnetic structure. (d) Radial cuts of SANS data showing the evolution of the ordering wave vector as a function of temperature taken at GP-SANS. Data in (b) and (c) are normalized to monitor counts, and data in (d) are scaled and vertically shifted for clarity.
} \label{f1}
\end{figure}

To confirm the accuracy and predictions of our model, a single crystal of U$_3$Al$_2$Ge$_3$ (tetragonal structure $I4$ (No. 79 in the International Tables for Crystallography), $a=7.769~\AA$, $c=11.036~\AA$, cf. Fig. \ref{f1}(a), derivative of ordered antitype-Cr$_5$B$_3$) was prepared by the Czochralski technique in a tri-arc furnace with a continuously purified Argon atmosphere ($< 10^{-12}~\mathrm{ppm}~\mathrm{O}_2$). The sample was characterized using X-ray Laue backscattering and via magnetic susceptibility measurements performed in a Quantum Design Magnetic Property Measurement System (MPMS). As shown in Fig. \ref{f0}, magnetic susceptibility data, both taken on warming after zero-field cooling, with no changes between runs, reveal typical ferromagnetic behavior with $T_{\mathrm{c}}=48~\mathrm{K}$, with a fully saturated susceptibility below 30~K, The susceptibility also exhibits an additional anomaly at 63~K corresponding to the onset temperature of the incommensurate magnetic state.\cite{rogl_magnetic_1999} For the small-angle neutron scattering (SANS) measurements, a 1~g piece of the single crystal was orientated such that [100] was along the beam and [001] was in the scattering plane, allowing to access the magnetic ordering wave vector $\mathbf{Q} = [0,0,\delta]$. SANS measurements were performed at the NG7-SANS beam line at the NIST Center for Neutron Research and the GP-SANS beam line at the High-Flux Isotope Reactor (HFIR) at Oak Ridge National Laboratory. SANS data were collected by rocking the sample $\pm10^{\circ}$ about the vertical axis, with an incident neutron wavelength of $\lambda=6~\AA$. 

\section{Experimental Results}
\label{results}
In the following section we present the results of the SANS measurements. All SANS data shown were obtained by angle integrating monitor-normalized SANS detector intensities. In order to obtain the magnetic intensity as a function of momentum transfer $q$, the magnetic scattering was azimuthally averaged in the FM phase, whereas in the incommensurate phase radial cuts through the two magnetic satellite positions at $\mathbf{Q}=[0,0,\delta]$ were performed.

Starting in the low-temperature FM phase, $T < 48$~K we only observe strong scattering near the direct beam and no characteristic magnetic ordering wave vectors, confirming the FM state. Furthermore, scattering near the direct beam approximately follows Porod's Law  $I\propto S q^{-4}$,\cite{porod_1951a,porod_1951b} as shown in Fig. \ref{f1}(b), which indicates the presence of large ferromagnetic domains with specific surface area $S = 0.251(7)~\mathrm{nm}^{2}$, from least-squares fitting.

Increasing temperature above 48~K we observe the emergence of strong magnetic scattering at $\mathbf{Q}=[0,0,\delta]$, indicating the entrance of the incommensurate phase. Although the original neutron powder diffraction study indicated this phase to be a sinusoidally modulated magnetic phase, from representational analysis of the crystallographic lattice structure determined in Ref. \onlinecite{rogl_magnetic_1999}, as shown in Fig. \ref{f1}(a), and the magnetic ordering wave vector $\mathbf{Q}=[0,0,\delta]$, both helical and sinusoidal magnetic ordering are allowed by symmetry. To confirm the magnetic structure, we therefore performed spin-polarized SANS measurements at the NG7-SANS instrument at $T=56$ K, with the polarization pointed along the [001] direction, and the detector positioned 4~m from the sample. In the case where the incident polarization is aligned along $q$, \textit{i.e.} $P_0 \parallel q$, the magnetic cross-section is given by
\begin{align}\label{magcrosssec}
{\sigma_{\pm}}&={\vert\bm{M}_{\perp}\vert^2\pm P_{0}C}.
\end{align}
Here $\bm{M}_{\perp}$ is the magnetic interaction vector defined as $\mathbf{M_{\perp}}=\hat{\mathbf{Q}}\times(\mathbf{\rho}(\mathbf{Q})\times\hat{\mathbf{Q}})$, where $\bf{\rho}(\mathbf{Q})=-2\mu_B\int\mathbf{\rho}(\mathbf{r})\exp(i\bf{Q}\cdot\mathbf{r})d\mathbf{r}$ is the Fourier transform of the magnetization density $\mathbf{\rho}(\bf{r})$ of the investigated sample, and $\hat{\mathbf{Q}}$ is a unit vector parallel to the scattering vector $\mathbf{Q}$. The associated coordinate frame is defined to have $x$ parallel to $\mathbf{Q}$, $z$ perpendicular to the scattering plane and $y$ completing the right-handed set. We note that the term $C=2\Im(M_{\perp y}^\ast\cdot M_{\perp z})$ is only non-zero for magnetic structures that display chirality, such as magnetic spirals, and is therefore denoted as the chiral term.\cite{janoschek_2007} Because, $(\sigma_{+} - \sigma_{-}) / P_0 = 2C$, we would expect that for a \emph{single} peak the difference between the cross section obtained with incident neutrons polarized parallel ($\sigma_{+}$) and antiparallel ($\sigma_{-}$) with respect to $q$ would give rise to a non-zero chiral term $C$, only if the magnetic structure had spiral order. This is contrary to our observation, where such a subtraction results in no residual intensity, within the error bars (cf. Fig. \ref{f1}(c)), confirming that the magnetic structure is sinusoidal, as surmised originally in Ref. \onlinecite{rogl_magnetic_1999}.

\begin{figure}[t]
\includegraphics[width=0.99\columnwidth]{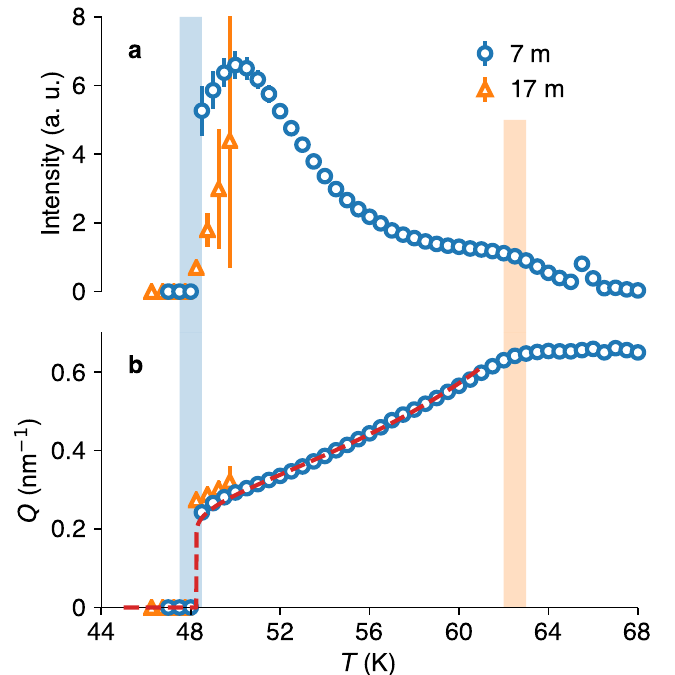}
\caption{(Color Online) Results of analysis of the SANS data: (a) the integrated intensity and (b) wave vector of the Bragg peaks, obtained by fitting Gaussian functions to radial cuts along the momentum transfer $q$, like those shown in Fig. \ref{f1}(d). Different symbols represent unique configurations: blue circles were obtained using a 7m detector distance, the orange triangles using a 17m detector distance. Error bars represent the errors obtained from least-squares fitting. The red dashed line in panel (b) represents a fit to the logarithmic singularity expression, $Q \propto -1/\ln(T-T_{\mathrm{c}})$.
} \label{f2}
\end{figure}

Further increase in temperature reveals a pronounced change, by a factor of three, in the magnetic ordering wave vector, as illustrated by the select SANS data for $49~\mathrm{K} \leq T \leq 67~\mathrm{K}$ shown in Fig. \ref{f1}(d), which have been scaled and shifted to illustrate the drastic change in the magnetic ordering wave vector as a function of temperature. In Fig. \ref{f2} we display the results of analysis of the angle-integrated SANS data for each temperature, taken at the GP-SANS instrument. For the temperature-dependent analysis we use data measured in two configurations: (1) with the detector at 7~m from the sample, which provides a momentum range of $0.1 \lesssim q \lesssim 0.8~\mathrm{nm}^{-1}$, and (2) with the detector 17~m from the sample, which provides a momentum range of $0.025 \lesssim q \lesssim 0.3~\mathrm{nm}^{-1}$. In the 7~m configuration, we can clearly see that the propagation vector $Q$ changes abruptly as a function of temperature at $T_{\mathrm{c}}\sim48$~K, while at 63~K the $Q$ saturates, as shown by the circle symbols in Fig. \ref{f2}(b). The 17~m detector configuration, due to its lower momentum range which allows us to observe scattering that would have been lost in the direct beam signal in the 7~m configuration, reveals that there is indeed a quasi-discontinuous change in the wave vector at $T_{\mathrm{c}}$. It is noteworthy that the intensity continues to increase slightly above the transition. This situation may arise in the case where the data were taken upon warming, as we have done here; since the transition from the FM to the modulated phase is first order, we may expect some hysteresis, \textit{i.e.} it is possible for ``droplets'' of the FM phase to persist above $T_{\mathrm{c}}$.\cite{PhysRevB.87.134407} This would therefore result in the appearance of the satellite peaks associated with the modulated phase immediately above $T_{\mathrm{c}}$, but with reduced integrated intensities. As the FM droplets shrink, the intensity of the satellite peaks will gradually increase, reaching a maximum at the temperature where the entire system is within the modulated phase, here $T\sim50$~K. It should also be noted here that in the 17~m configuration, the ordering wave vector of the sinusoidal phase lies at the limits of the detector range, resulting in substantial error in the intensity, as illustrated by the error bars in Fig. \ref{f2}(b). Furthermore, we find that the wave vector is well described by the logarithmic function $-1/\ln(T-T_{\mathrm{c}})$ (cf. Eq. \ref{eq15cc}), as illustrated by the dashed line in Fig. \ref{f2}(b). This function directly originates from the ANNNI models, as we discussed above in Section \ref{model}. This quasi-discontinuity in $Q$ indicates a first-order transition between the sinusoidal and FM phases, supporting the assertion of a first-order transition predicted by our model.

In contrast, above $T_{\mathrm{N}}=63$~K we observe no discrete peaks on the detector, but instead a broad ring of scattering which gradually decreases in intensity up to $\sim$67~K, above which it disappears below the background completely. This behavior, along with the magnetization data in Fig. \ref{f0}, strongly supports a second-order transition between the paramagnetic (PM) and sinusoidal phase.

\section{Discussion and summary}
\label{discussion}
It is found that the lattice constants change in the sinusoidally modulated phase, indicating the existence of magneto-elastic coupling.\cite{rogl_magnetic_1999} Therefore, the measured change of $Q$ by neutron scattering has two contributions, with one being from the lattice expansion/shrinkage, and the other from the magnetic competing interaction. The former contribution is negligible compared to the latter one according to the experiments.\cite{rogl_magnetic_1999}
  
One key feature of the ANNNI framework on a lattice model is the appearance of the devil's staircase, where $Q$ varies quasi-continuously with temperature. In certain temperature window, $Q$ is fixed. Outside the temperature window, $Q$ then jumps to another $Q$ through the proliferation of magnetic solitons. \cite{PhysRevB.21.5297} This behavior is not captured by the model in the continuum limit in Eq. \eqref{eq1}. For $Q=0.76~\mathrm{nm}^{-1}$, the $Q$ steps are extremely narrow. In experiments with finite resolution, $Q$ varies nearly continuously with temperature, as depicted in Fig. \ref{f2}. 

True realizations of the ANNNI model in magnets is rare, particularly in metals. In fact many of the potential examples of ANNNI-like materials require notable modification to the ANNNI model to explain observed behavior. The best known example is semimetallic CeSb,\cite{fischer_1978} the magnetization of which features the famous ``devil's staircase'', discrete sharp steps in the magnetization within the incommensurate phase.\cite{bak_commensurate_1982} However, the ANNNI model notably does not capture additional experimentally observed Bragg reflections in CeSb, requiring a modification of the ANNNI model.\cite{meier_1978,pokrovskii_1982} A later study of CeBi also required modification of the ANNNI model, \textit{i.e.} an additional competing exchange coupling.\cite{uimin_1982} More recently, the phase diagram of metallic TmB$_4$, which features a
fractional plateau in magnetization, has been described in terms of the ANNNI model, but again, two additional exchange terms were required as modification of the model.\cite{wierschem_2015}

Several other modern examples of potential ANNNI compounds do exist, however. In U(Ru$_{1-x}$Rh$_x$)$_2$Si$_2$, powder neutron diffraction observed multiple $Q$ states as a function of temperature, suggested by the authors to perhaps be related to the ANNNI model.\cite{kawarazaki_1994} A recent study has reported a semimetallic compound closely related to CeSb, CeSbSe which was found to feature discrete steps in magnetization and resistivity, which have been suggested to arise due to underlying ANNNI type magnetic interactions and the associated existence of a devil's staircase.\cite{PhysRevB.96.014421} The heavy fermion metal CeRhIn$_5$, which exhibits incommensurate helical and elliptical magnetic ordering that transition to a commensurate order with magnetic field,\cite{raymond_2007} has been suggested as a candidate.\cite{das_2015,fobes_2017}

In summary, we have studied the magnetic order in $f$-electron compound U$_3$Al$_2$Ge$_3$, finding it to be well described by the three-dimensional Axial Next-Nearest Neighbor Ising model. Our experimental results determine the transition between the paramagnetic and sinusoidal phases to be second order, and the lower transition between the sinusoidal and ferromagnetic phases to be first order, which arises naturally from our model. Furthermore, we have confirmed the logarithmic singularity in the sinusoidal ordering wave vector at the boundary to the FM phase, an essential feature of the ANNNI model. These results demonstrate that U$_3$Al$_2$Ge$_3$ represents a prototypical material for investigating ANNNI physics.

\section{Acknowledgements}
Research at LANL was supported by the LANL Directed Research and Development program (neutron scattering) and the U.S. Department of Energy, Office of Basic Energy Sciences, Division of Materials Sciences and Engineering under the project `Complex Electronic Materials' (material synthesis and characterization). A portion of this research used resources at the High Flux Isotope Reactor, a DOE Office of Science User Facility operated by the Oak Ridge National Laboratory. We acknowledge the support of the National Institute of Standards and Technology, U.S. Department of Commerce, in providing the neutron research facilities used in this work. Access to the polarized NG7SANS was provided by the Center for High Resolution Neutron Scattering, a partnership between the National Institute of Standards and Technology and the National Science Foundation under Agreement No. DMR-1508249. We thank Cedric Gagnon and Jeff Krzywon for technical support during the NG7SANS measurements. Data reduction and analysis was performed with the use of the GRASP and NeutronPy neutron scattering analysis software packages.


%

\end{document}